\documentclass[10pt]{article}

\usepackage[utf8]{inputenc}
\usepackage[english]{babel}
\usepackage{amsmath,amsfonts,amssymb}
\usepackage{graphicx}
\usepackage{float}
\usepackage{booktabs}
\usepackage{url}
\usepackage{hyperref}

\usepackage[utf8]{inputenc}
\usepackage[english]{babel}
\usepackage{amsmath,amsfonts,amssymb}
\usepackage{graphicx}
\usepackage{float}
\usepackage{booktabs}
\usepackage{url}
\usepackage{hyperref}
\usepackage[round]{natbib}

\usepackage[letterpaper,margin=1in]{geometry}

\title{100-Day Analysis of USD/IDR Exchange Rate Dynamics Around the 2025 U.S. Presidential Inauguration}

\author{
Sandy H. S. Herho$^{1,*}$, Siti N. Kaban$^{2}$, Cahya Nugraha$^{3}$ \\
\\
\small $^{1}$Department of Earth and Planetary Sciences, University of California, Riverside, CA, USA 92521\\
\small $^{2}$Financial Engineering Program, WorldQuant University, Washington, D.C., USA 20002\\
\small $^{3}$Al Mumtaaz Islamic Charitable Foundation, Karawang, West Java, Indonesia 41361\\
\\
\small $^{*}$Corresponding author: sandy.herho@email.ucr.edu
}

\date{}

\begin{document}

\maketitle

\begin{abstract}
Using a 100-day symmetric window around the January 2025 U.S. presidential inauguration, non-parametric statistical methods with bootstrap resampling (10,000 iterations) analyze distributional properties and anomalies. Results indicate a statistically significant 3.61\% Indonesian rupiah depreciation post-inauguration, with a large effect size (Cliff's Delta $= -0.9224$, CI: $[-0.9727, -0.8571]$). Central tendency shifted markedly, yet volatility remained stable (variance ratio $= 0.9061$, $p = 0.504$). Four significant anomalies exhibiting temporal clustering are detected. These findings provide quantitative evidence of political transition effects on emerging market currencies, highlighting implications for monetary policy and currency risk management.
\end{abstract}


\section{Introduction}

Political transitions in major economies generate significant ripple effects across global financial markets, with emerging market currencies often experiencing pronounced volatility during these periods \citep{Block2003, Mei2004, Tabash2024}. Exchange rates, as key economic indicators, reflect market sentiments and expectations regarding future policy directions. The inauguration of Donald Trump for his second presidential term on January 20, 2025, presents a valuable opportunity to examine how political transitions in the United States affect currency dynamics in emerging economies, particularly Indonesia.

This study employs a symmetric 100-day window before and after the presidential inauguration, a timeframe that captures both anticipatory market reactions and post-transition policy implementation effects. The 100-day period represents a critical interval wherein market participants process pre-election signals, inauguration events, and initial policy actions, while providing sufficient statistical power to detect meaningful patterns without introducing excessive noise from unrelated economic factors. Prior studies on political transition effects suggests that market adjustments typically manifest most prominently within this temporal boundary \citep{Eshbaugh2005, Fauvelle2013, Selmi2020}, with anticipatory positioning occurring in the pre-inauguration phase and policy uncertainty resolution developing in the post-inauguration phase.

Indonesia's economic standing in Southeast Asia and its integration with global markets make the Indonesian Rupiah (IDR) a compelling case study for analyzing the effects of external political developments. The country maintains significant external financing requirements and a relatively thin domestic financial market, which potentially amplifies the transmission of global shocks to local financial conditions. Despite Indonesia's strong macroeconomic fundamentals, with GDP growth consistently above five percent in recent years \citep{Resosudarmo2018}, the country's financial markets exhibit sensitivity to global risk factors.

Political transitions influence exchange rates through multiple interconnected channels. The policy uncertainty mechanism suggests that ambiguity surrounding future policy directions leads to risk premiums in currency markets. For Indonesia, this takes on particular significance due to the country's integration with global financial markets and reliance on U.S. Dollar (USD)-denominated trade. The interest rate differential mechanism explains how political transitions alter expectations about future central bank policies, thereby affecting relative returns on financial assets across countries. Additionally, global risk sentiment can trigger significant reallocation of capital across markets, with investors typically reducing exposure to emerging market assets during periods of heightened uncertainty.

Recent market developments indicate potential implications of Trump's second administration for emerging markets. His policy platform emphasizing tariffs, trade renegotiations, and domestic manufacturing could alter global trade patterns and capital flows \citep{Rosenberger2024, Chohan2025}. Indonesia, as an export-oriented economy with significant trade connections to both the United States and China, faces potential exposure to these policy shifts. Bank Indonesia has acknowledged that political transitions necessitate vigilance in currency management, with the central bank consistently prioritizing IDR stability during periods of political uncertainty \citep{Reuters2025BI}.

This study examines three critical hypotheses within the 100-day window framework. First, we investigate whether the USD/IDR exchange rate exhibits different distributional characteristics before and after the presidential inauguration. Second, we analyze whether exchange rate volatility differs significantly between pre-inauguration and post-inauguration periods. Third, we identify anomalous exchange rate behaviors potentially attributable to specific policy announcements or market reactions during the transition period. Understanding these dynamics carries significant implications for monetary policy formulation, foreign exchange risk management strategies, and international investment decisions. The findings contribute to the growing literature on political risk pricing in currency markets and the international transmission of policy uncertainty from core economies to emerging markets.

\section{Literature Review}

The relationship between political transitions and exchange rate dynamics represents one of the most consequential areas of international macroeconomics, particularly for emerging market economies like Indonesia. Political transitions in major economies, especially the United States, generate significant ripple effects across global financial markets, with emerging market currencies often experiencing pronounced volatility during these periods \citep{Bernhard2002}. The theoretical foundations of this relationship trace back to the open-economy macroeconomic models developed by \citet{Mundell1963} and \citet{Fleming1962}, though contemporary understanding has evolved to incorporate a sophisticated blend of traditional economic mechanisms and political economy considerations.

Exchange rate responses to political transitions operate through multiple interconnected channels that have been extensively documented in the literature. The policy uncertainty mechanism, formalized by \citet{Baker2016}, posits that ambiguity surrounding future policy directions during transitions leads to risk premiums in currency markets. This uncertainty manifests in expectations about future monetary, fiscal, and trade policies, which directly influence capital flows and, consequently, exchange rates. For Indonesia, this channel takes on particular significance due to the country's integration with global financial markets and reliance on dollar-denominated trade. According to \citet{Reuters2025Bold}, Bank Indonesia has explicitly acknowledged that political transitions necessitate vigilance in currency management, with the central bank consistently prioritizing rupiah stability during periods of political uncertainty.

The interest rate differential mechanism explains exchange rate dynamics during transitions through the relationship between relative returns on financial assets across countries, as established in the seminal work of \citet{Fama1984}. Political transitions alter expectations about future central bank policies, thereby affecting interest rate differentials and, consequently, exchange rates. \citet{Reuters2023fall} documented that Bank Indonesia has explicitly incorporated this mechanism into its policy framework, employing a multi-instrument approach that includes policy rate adjustments, market intervention, and macroprudential measures. In October 2023, the central bank delivered a surprise 25 basis point rate hike primarily to support the rupiah, which had fallen to its lowest level since 2020 amid rising global risk aversion and revised expectations about U.S. monetary policy easing.

Global risk sentiment forms a third critical channel through which political transitions affect emerging market currencies, as articulated in the influential study of \citet{Rey2015} on the "global financial cycle." Major political shifts in the U.S. can trigger significant reallocation of capital across global markets, with investors typically reducing exposure to emerging market assets during periods of heightened uncertainty. The \citet{FT2016} reported that emerging market currencies, including the IDR, experienced significant pressure during the 2016 U.S. presidential transition, with portfolio investors reducing exposure to these markets amid uncertainty about future U.S. trade and foreign policies. \citet{Kalemli2019} quantified this effect, finding that emerging markets typically experience capital flow reversals averaging 1.2\% of GDP during U.S. presidential transitions.

The IDR's historical behavior during U.S. presidential transitions reveals consistent patterns of volatility that inform expectations for current and future transitions. \citet{Basri2020} documented that during the 2016-2017 transition, the USD/IDR exchange rate exhibited substantial depreciation followed by a gradual recovery as policy uncertainty diminished. Similarly, \citet{Warjiyo2019} analyzed the 2008-2009 transition, which coincided with significant IDR volatility, though this was confounded by the global financial crisis. These historical patterns provide a framework for analyzing potential outcomes during political transitions, with early indicators from current market developments suggesting similar dynamics. According to \citet{TradingEconomics2023}, the Indonesian IDR experienced notable depreciation amid heightened global uncertainty, with the central bank maintaining vigilant intervention in the currency markets.

Indonesia's specific vulnerabilities to external political shocks stem from several structural economic characteristics despite its strong fundamentals, as analyzed by the \citet{IMF2023}. The country maintains significant external financing requirements and a relatively thin domestic financial market, which amplifies the transmission of global shocks to local financial conditions. While Indonesia's macroeconomic fundamentals remain strong, with GDP growth consistently above moderate levels in recent years, the country's financial markets exhibit heightened sensitivity to global risk factors, particularly during periods of political transition. This sensitivity manifests in both exchange rate volatility and capital flow dynamics, with portfolio investment flows demonstrating pronounced reactions to U.S. political developments as documented by \citet{BankIndonesia2023} in their quarterly reports.

The current political and economic landscape presents unique challenges for the IDR. The \citet{WorldBank2023} reports that Indonesia faces a double transition effect, with domestic political transitions occurring in proximity to global political transitions, potentially creating compound effects for currency markets. This domestic political transition has its own implications for economic policy uncertainty, which may compound the effects of external political developments. Under Indonesia's current leadership, the country faces important policy decisions regarding fiscal consolidation, investment attraction, and monetary policy independence, all of which have implications for exchange rate stability during periods of global political transition.

Recent data underscores the IDR's vulnerability to transition periods. According to \citet{Reuters2023fall}, the IDR fell to multi-year lows against the dollar in late 2023, with Bank Indonesia delivering a surprise rate hike to anchor the currency. This depreciation occurred as markets priced in potential impacts of global trade friction and monetary policy shifts in advanced economies. These market reactions align with theoretical predictions from \citet{Pastor2013}, whose model implies that political uncertainty commands a risk premium whose magnitude is larger during weak economic conditions.

Building on these empirical observations, recent theoretical advances have refined our understanding of how political transitions affect emerging market currencies. \citet{Bailey1995} established that political risk is distinctly priced in emerging market currencies, creating systematic variation that cannot be diversified away. This framework has been extended by \citet{Filippou2018}, who demonstrate that measures of relative political stability significantly explain time variation in currency risk premia. The transmission mechanisms identified in this literature operate through multiple channels: policy uncertainty affecting risk premiums, interest rate differential adjustments, and shifts in global risk sentiment that trigger capital flow reversals.

The methodological evolution in analyzing these effects has been substantial. Contemporary approaches, as reviewed by \citet{Eichengreen2018}, increasingly employ nonparametric techniques and robust statistical methods that accommodate the non-normal distributions typical of exchange rate data during turbulent periods. This methodological sophistication is particularly important given the documented deterioration in market liquidity during political transitions \citep{Menkhoff2016}, which amplifies price movements and contributes to exchange rate overshooting in thin markets like the Indonesian IDR.

Indonesia's policy response framework has adapted to these challenges through institutional innovations. Bank Indonesia has developed a multi-instrument approach combining traditional interest rate adjustments with foreign exchange interventions and macroprudential measures \citep{Warjiyo2019}. This comprehensive toolkit reflects recognition that political transition effects require coordinated policy responses across multiple dimensions. However, as \citet{Gourinchas2011} emphasize, Indonesia's position in the "international monetary system's periphery" means that even sophisticated policy frameworks cannot fully insulate the IDR from U.S. political transitions, necessitating continued vigilance and adaptive policy responses.

\section{Method}
\subsection{Data Collection and Preprocessing}

The daily USD/IDR exchange rate data were collected through the \texttt{yfinance} Python library \citep{Ranaraja2022}, which provided programmatic access to Yahoo Finance historical market data. We examined a symmetric window of 100 calendar days before and after the presidential inauguration (October 14, 2024 to April 29, 2025).

The raw data were structured into three distinct analytical segments:
\begin{align}
\mathbf{X}_{\text{pre}} &= \{x_t : t \in [t_0, t_{\text{inaug}}-1]\}, \\
\mathbf{X}_{\text{post}} &= \{x_t : t \in [t_{\text{inaug}}, t_{\text{end}}]\}, \\
\mathbf{X}_{\text{full}} &= \mathbf{X}_{\text{pre}} \cup \mathbf{X}_{\text{post}},
\end{align}
where $x_t$ represented the USD/IDR exchange rate at time $t$, $t_0$ denoted the starting date (October 14, 2024), $t_{\text{inaug}}$ represented the inauguration date (January 20, 2025), and $t_{\text{end}}$ denoted the ending date (April 29, 2025).

Missing values were identified using the approach in Equation \eqref{eq:missing}, where $\mathbf{1}$ represented the indicator function:
\begin{equation}
\mathcal{M} = \{t : \mathbf{1}(x_t = \text{NaN}) = 1\}.
\label{eq:missing}
\end{equation}
Rather than imputing missing values, we excluded them from calculations. The data were temporally aligned, accounting for differences in trading calendars between the US and Indonesia.

For outlier identification, we employed robust statistical measures. Following \citet{Rousseeuw2011}, we calculated modified $Z$-scores:
\begin{equation}
Z_i = \frac{0.6745(x_i - \tilde{x})}{\text{MAD}},
\label{eq:modz}
\end{equation}
where $\tilde{x}$ was the median and MAD was the median absolute deviation. Observations were flagged for further examination when $|Z_i| > 3.5$, though these observations were retained in the analysis while employing robust statistical methods.

The dataset properties were characterized using the \texttt{pandas} \citep{McKinney2010}, \texttt{numpy} \citep{Harris2020}, and \texttt{scipy} \citep{Virtanen2020} libraries. We calculated robust measures of central tendency and dispersion:

\begin{align}
\bar{x}_{\alpha} &= \frac{1}{n(1-2\alpha)}\sum_{i=k+1}^{n-k}x_{(i)} \quad \text{where } k = \lfloor \alpha n \rfloor, \label{eq:trimean}\\
\text{MAD} &= \text{median}(|x_i - \tilde{x}|), \label{eq:mad}\\
\text{IQR} &= Q_3 - Q_1, \label{eq:iqr}
\end{align}

where $\bar{x}_{\alpha}$ represented the $\alpha$-trimmed mean, $x_{(i)}$ represented the $i$-th order statistic, MAD was the median absolute deviation, and IQR was the interquartile range. We implemented 10\% and 20\% trimmed means ($\alpha = 0.1$ and $\alpha = 0.2$, respectively) to assess the impact of extreme observations on central tendency estimates \citep{Wilcox2012}.

Additionally, we characterized higher-order moments of the exchange rate distributions:

\begin{align}
\gamma_1 &= \frac{\frac{1}{n}\sum_{i=1}^{n}(x_i-\bar{x})^3}{\left(\frac{1}{n}\sum_{i=1}^{n}(x_i-\bar{x})^2\right)^{3/2}}, \label{eq:skew}\\
\gamma_2 &= \frac{\frac{1}{n}\sum_{i=1}^{n}(x_i-\bar{x})^4}{\left(\frac{1}{n}\sum_{i=1}^{n}(x_i-\bar{x})^2\right)^{2}} - 3, \label{eq:kurt}
\end{align}

where $\gamma_1$ represented skewness and $\gamma_2$ represented excess kurtosis.

\subsection{Normality Assessment Framework}

To determine the appropriate statistical methodology, we implemented a comprehensive normality assessment framework. We evaluated the distributional characteristics of the exchange rate data across all analytical segments ($\mathbf{X}_{\text{pre}}$, $\mathbf{X}_{\text{post}}$, and $\mathbf{X}_{\text{full}}$) using a suite of complementary normality tests.

The Shapiro-Wilk test \citep{Shapiro1965} evaluated the null hypothesis $H_0$ that the sample $\mathbf{X}$ came from a normally distributed population. The test statistic $W$ was computed as:

\begin{equation}
W = \frac{\left(\sum_{i=1}^{n}{a_i x_{(i)}}\right)^2}{\sum_{i=1}^{n}{(x_i - \bar{x})^2}},
\label{eq:shapiro}
\end{equation}

where $x_{(i)}$ were the ordered sample values, $\bar{x}$ was the sample mean, and the constants $a_i$ were derived from the means, variances, and covariances of the order statistics from a standard normal distribution.

The Anderson-Darling test \citep{Anderson1952}, which placed greater emphasis on distribution tails, was implemented as:

\begin{equation}
A^2 = -n - \frac{1}{n}\sum_{i=1}^{n}{(2i-1)[\ln\Phi(z_{(i)}) + \ln(1-\Phi(z_{(n+1-i)}))]},
\label{eq:anderson}
\end{equation}

where $\Phi$ represented the cumulative distribution function of the standard normal distribution, and $z_{(i)} = \frac{x_{(i)}-\bar{x}}{s}$ were the standardized ordered observations.

The Jarque-Bera test \citep{Jarque1987} evaluated normality through skewness and kurtosis:

\begin{equation}
JB = \frac{n}{6}\left(\gamma_1^2 + \frac{(\gamma_2)^2}{4}\right),
\label{eq:jarquebera}
\end{equation}

which followed an asymptotic $\chi^2$ distribution with two degrees of freedom under the null hypothesis of normality.

For additional robustness, we implemented the D'Agostino-Pearson test \citep{DAgostino1973}, which combined skewness and kurtosis into an omnibus test:

\begin{equation}
K^2 = Z^2(\gamma_1) + Z^2(\gamma_2),
\label{eq:dagostino}
\end{equation}

where $Z(\gamma_1)$ and $Z(\gamma_2)$ were transformations of the sample skewness and kurtosis statistics to approximate normality.

Finally, we implemented the Lilliefors test \citep{Lilliefors1967}, a modified Kolmogorov-Smirnov test for normality when population parameters were estimated from the sample:

\begin{equation}
D = \max\left\{\max_{1 \leq i \leq n}\left\{\frac{i}{n} - \Phi\left(\frac{x_{(i)}-\bar{x}}{s}\right)\right\}, \max_{1 \leq i \leq n}\left\{\Phi\left(\frac{x_{(i)}-\bar{x}}{s}\right) - \frac{i-1}{n}\right\}\right\},
\label{eq:lilliefors}
\end{equation}

where critical values were adjusted for parameter estimation effects.

We implemented a decision rule where rejection by at least three tests at $\alpha=0.05$ indicated significant departure from normality.

To complement these formal tests, we employed kernel density estimation (KDE) using the Gaussian kernel with bandwidth selection following \citet{Silverman1986}:

\begin{equation}
\hat{f}_h(x) = \frac{1}{nh}\sum_{i=1}^{n}K\left(\frac{x-x_i}{h}\right),
\label{eq:kde}
\end{equation}

where $K(u) = \frac{1}{\sqrt{2\pi}}e^{-\frac{u^2}{2}}$ was the kernel function and the bandwidth parameter $h$ was calculated as:

\begin{equation}
h = 0.9 \times \min\left\{\hat{\sigma}, \frac{\text{IQR}}{1.34}\right\} \times n^{-1/5},
\label{eq:bandwidth}
\end{equation}

with $\hat{\sigma}$ representing the sample standard deviation.

We further characterized the shape of distributions using L-moments \citep{Hosking1990}, which provided more robust measures of skewness ($\tau_3$) and kurtosis ($\tau_4$) than conventional moments:

\begin{align}
\tau_3 &= \frac{L_3}{L_2}, \label{eq:lskew}\\
\tau_4 &= \frac{L_4}{L_2}, \label{eq:lkurt}
\end{align}

where $L_r$ represented the $r$-th L-moment, defined through probability weighted moments:

\begin{equation}
L_r = \sum_{j=0}^{r-1}{(-1)^{r-1-j}\binom{r-1}{j}\binom{r-1+j}{j}\beta_j},
\label{eq:lmoment}
\end{equation}

with $\beta_j = \frac{1}{n}\sum_{i=1}^{n}{\binom{i-1}{j}/\binom{n-1}{j}}x_{(i)}$.

\subsection{Robust Non-parametric Statistical Framework}

Building upon the normality assessment outcomes, we implemented a comprehensive non-parametric statistical framework. The core of our analytical approach incorporated bootstrap resampling with multiple non-parametric tests. Following \citet{Efron1994}, we implemented the bootstrap methodology with $B = 10,000$ iterations to generate empirical sampling distributions for each test statistic.

For a given bootstrap sample $b$, we drew with replacement from the original dataset:
\begin{equation}
\mathbf{X}^*_b = \{x^*_{b,1}, x^*_{b,2}, \ldots, x^*_{b,n}\},
\label{eq:bootstrap}
\end{equation}

where each $x^*_{b,i}$ was randomly sampled with replacement from the original observations. This procedure was repeated independently for both pre-inauguration and post-inauguration periods.

For each bootstrap iteration, we calculated the test statistic $\hat{\theta}^*_b$ and derived the empirical $p$-value as:
\begin{equation}
\hat{p} = \frac{1}{B}\sum_{b=1}^{B}\mathbf{1}(\hat{\theta}^*_b \geq \hat{\theta}),
\label{eq:pval}
\end{equation}

where $\mathbf{1}(\cdot)$ was the indicator function. We implemented the rejection ratio approach to determine robust significance, defined as:
\begin{equation}
\text{RR} = \frac{1}{B}\sum_{b=1}^{B}\mathbf{1}(p^*_b < \alpha),
\label{eq:rr}
\end{equation}

where $p^*_b$ was the $p$-value for bootstrap sample $b$ and $\alpha = 0.05$ was the significance level. A result was considered robustly significant when $\text{RR} > 0.8$.

Our framework incorporated four complementary non-parametric tests:

First, we implemented the Brown-Forsythe test \citep{Brown1974} to evaluate variance homogeneity:
\begin{equation}
F_{BF} = \frac{(N-k)\sum_{i=1}^{k}n_i(\bar{Z}_{i.} - \bar{Z}_{..})^2}{(k-1)\sum_{i=1}^{k}\sum_{j=1}^{n_i}(Z_{ij} - \bar{Z}_{i.})^2},
\label{eq:bf}
\end{equation}

where $Z_{ij} = |X_{ij} - \tilde{X}_i|$ represented the absolute deviation from group median $\tilde{X}_i$, $\bar{Z}_{i.}$ and $\bar{Z}_{..}$ were group and overall means of these deviations, $n_i$ was the sample size of group $i$, $k=2$ was the number of groups, and $N = n_1 + n_2$ was the total sample size.

Second, we calculated Cliff's Delta \citep{Cliff1993}, a robust non-parametric effect size measure:
\begin{equation}
\delta = \frac{\sum_{i=1}^{n_1}\sum_{j=1}^{n_2}\text{sgn}(x_{1i} - x_{2j})}{n_1 n_2},
\label{eq:cliff}
\end{equation}

where $\text{sgn}(\cdot)$ was the signum function. We interpreted $|\delta|$ values as: negligible ($<0.147$), small ($<0.33$), medium ($<0.474$), or large ($\geq0.474$).

Third, we implemented the Kolmogorov-Smirnov test \citep{Kolmogorov1933, Smirnov1948}, which evaluated the maximum difference between empirical cumulative distribution functions:
\begin{equation}
D_{KS} = \sup_x |F_1(x) - F_2(x)|,
\label{eq:ks}
\end{equation}

where $F_1(x)$ and $F_2(x)$ were the empirical cumulative distribution functions for pre-inauguration and post-inauguration periods.

Fourth, we applied the Mann-Whitney U test \citep{Mann1947}, which evaluated whether one distribution was stochastically greater than another:
\begin{equation}
U = n_1 n_2 + \frac{n_1(n_1+1)}{2} - R_1,
\label{eq:mwu}
\end{equation}

where $R_1$ was the sum of ranks in the first sample when both samples were combined and ranked.

\subsection{Anomaly Detection and Volatility Analysis}

To identify anomalous exchange rate behaviors, we implemented an ensemble anomaly detection framework. Our approach integrated multiple detection methods to mitigate algorithm-specific limitations.

The ensemble methodology combined three distinct algorithms: Isolation Forest, One-Class SVM, and a statistical approach based on robust outlier detection. Following \citet{Liu2008}, we implemented Isolation Forest with optimized hyperparameters:

\begin{equation}
s(x, n) = 2^{-\frac{E(h(x))}{c(n)}},
\label{eq:isoforest}
\end{equation}

where $s(x, n)$ represented the anomaly score for observation $x$ in a dataset of size $n$, $E(h(x))$ was the average path length for isolating observation $x$, and $c(n) = 2H(n-1) - \frac{2(n-1)}{n}$ was the average path length of unsuccessful search in a binary search tree. The algorithm employed $n_t = 300$ trees with subsampling size $\psi = 256$ and contamination parameter $\varepsilon = 0.05$.

For the One-Class SVM component, we implemented the methodology of \citet{Scholkopf2001} with a radial basis function (RBF) kernel:

\begin{equation}
f(x) = \text{sgn}\left(\sum_{i=1}^{n} \alpha_i K(x_i, x) - \rho\right),
\label{eq:ocsvm}
\end{equation}

where $\alpha_i$ were Lagrange multipliers, $K(x_i, x) = \exp(-\gamma \|x_i - x\|^2)$ was the RBF kernel with parameter $\gamma = \frac{1}{d \cdot \sigma^2_X}$, and $\rho$ was the bias term. We set the regularization parameter $\nu = 0.05$ to constrain the fraction of potential outliers.

The statistical approach employed robust outlier detection based on interquartile range:

\begin{equation}
\mathcal{O}(x_t) = 
\begin{cases}
-1, & \text{if } x_t < Q_1 - 1.5 \cdot \text{IQR} \text{ or } x_t > Q_3 + 1.5 \cdot \text{IQR} \\
1, & \text{otherwise}
\end{cases},
\label{eq:iqrout}
\end{equation}

We implemented a consensus-based approach to integrate these methods. For each observation $x_t$, we calculated a weighted ensemble score:

\begin{equation}
\Psi(x_t) = \omega_{\text{IF}} \cdot \Psi_{\text{IF}}(x_t) + \omega_{\text{OCSVM}} \cdot \Psi_{\text{OCSVM}}(x_t) + \omega_{\text{STAT}} \cdot \Psi_{\text{STAT}}(x_t),
\label{eq:ensemble}
\end{equation}

where $\omega_{\text{IF}} = 0.4$, $\omega_{\text{OCSVM}} = 0.4$, and $\omega_{\text{STAT}} = 0.2$ were the respective weights.

The final anomaly classification employed a voting mechanism:
\begin{equation}
\mathcal{A}(x_t) = 
\begin{cases}
-1, & \text{if } \sum_{m \in \mathcal{M}} \mathbf{1}(\mathcal{D}_m(x_t) = -1) \geq 2 \\
1, & \text{otherwise}
\end{cases},
\label{eq:anomvote}
\end{equation}
where $\mathcal{M} = \{\text{IF}, \text{OCSVM}, \text{STAT}\}$ represented the set of detection methods.

For volatility analysis, we examined volatility across multiple time scales through rolling window calculations:

\begin{equation}
\hat{\sigma}_t^2(w) = \frac{1}{w-1} \sum_{i=t-w+1}^{t} (r_i - \bar{r}_t(w))^2,
\label{eq:rollingvar}
\end{equation}
where $r_i = \ln \frac{x_i}{x_{i-1}}$ represented the logarithmic return at time $i$, and $\bar{r}_t(w) = \frac{1}{w} \sum_{i=t-w+1}^{t} r_i$ was the mean return over the window. We implemented this calculation for $w \in \{7, 14\}$ days.

We defined the bootstrap confidence interval for the variance ratio between post-inauguration and pre-inauguration periods:
\begin{equation}
\text{CI}_{1-\alpha}\left(\frac{\sigma^2_{\text{post}}}{\sigma^2_{\text{pre}}}\right) = \left[Q_{\alpha/2}\left(\frac{\hat{\sigma}^{2*}_{\text{post}}}{\hat{\sigma}^{2*}_{\text{pre}}}\right), Q_{1-\alpha/2}\left(\frac{\hat{\sigma}^{2*}_{\text{post}}}{\hat{\sigma}^{2*}_{\text{pre}}}\right)\right],
\label{eq:bootci}
\end{equation}
where $Q_p$ represented the $p$-th quantile of the bootstrap distribution.

\section{Result and Analysis}

\begin{figure}[H]
\centering
\includegraphics[width=\textwidth]{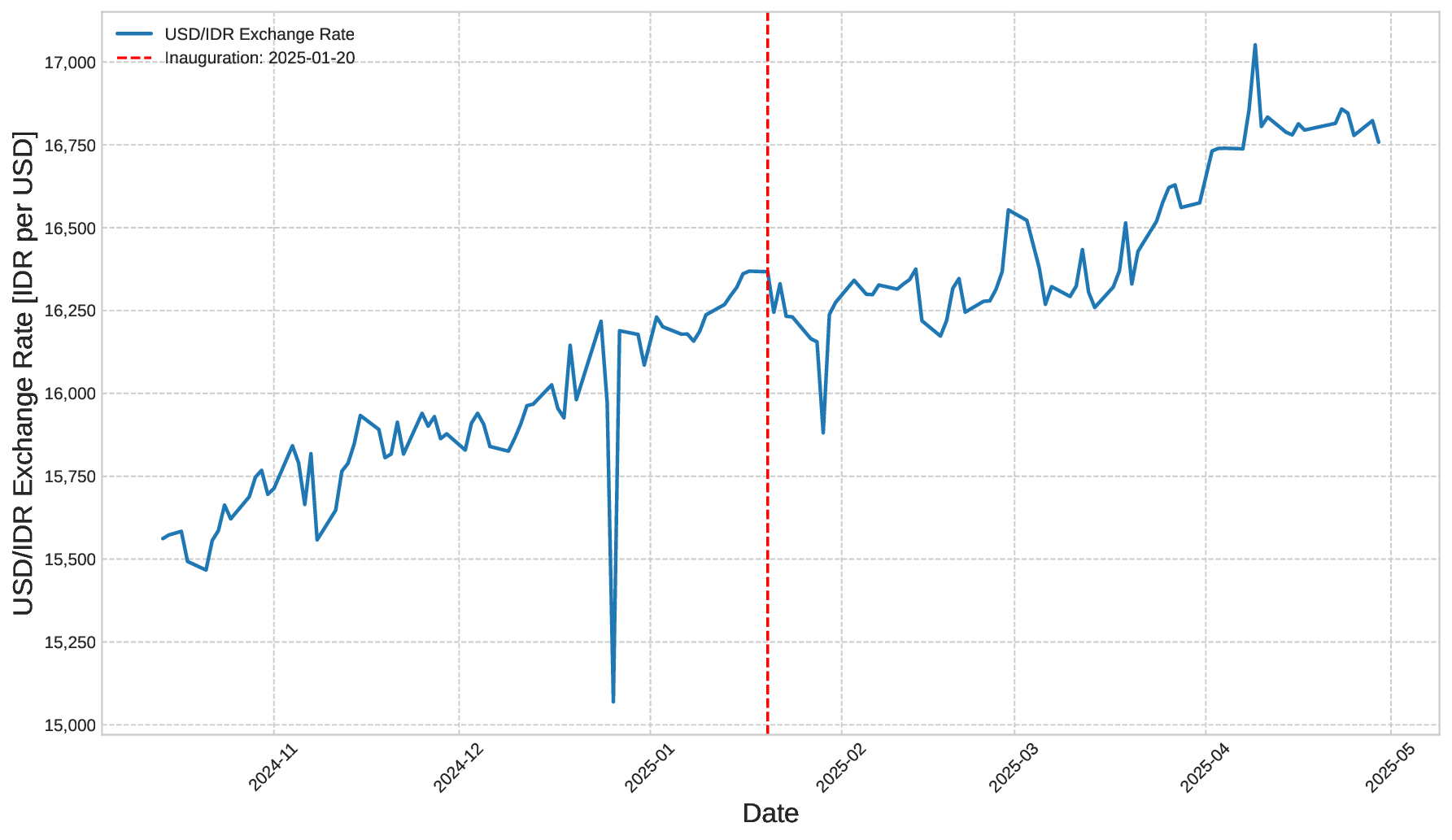}
\caption{Time series of USD/IDR exchange rate over the 100-day window before and after the January 20, 2025 presidential inauguration. The vertical red dashed line indicates the inauguration date.}
\label{fig:timeseries}
\end{figure}

Analysis of USD/IDR exchange rate data reveals substantial differences between pre-inauguration and post-inauguration periods. Figure \ref{fig:timeseries} presents the 100-day symmetric window time series, demonstrating a clear upward trend in exchange rate following the presidential inauguration. This pattern represents a statistically significant and economically meaningful shift in the USD/IDR exchange rate, with the rupiah experiencing a 3.61\% depreciation against the US dollar in the post-inauguration period.

Descriptive statistics presented in Table \ref{tab:descriptive} quantify this shift, with the mean USD/IDR exchange rate increasing from 15,892.01 IDR/USD (SD = 249.49) in the pre-inauguration period to 16,465.54 IDR/USD (SD = 237.51) in the post-inauguration period. The median values likewise increased from 15,891.30 to 16,367.25 IDR/USD. While standard deviations are similar between periods, the interquartile range (IQR) expanded from 338.60 to 413.20, suggesting a broader distribution of values after the inauguration. The direction and magnitude of these movements align with established theoretical frameworks of political uncertainty transmission to emerging market currencies.

\begin{table}[H]
\centering
\caption{Descriptive Statistics of USD/IDR Exchange Rate}
\begin{tabular}{lrrr}
\hline
\textbf{Metric} & \textbf{Pre-Inauguration} & \textbf{Post-Inauguration} & \textbf{Entire Period} \\
\hline
Count & 69 & 70 & 139 \\
Mean & 15,892.01 & 16,465.54 & 16,180.84 \\
Median & 15,891.30 & 16,367.25 & 16,230.90 \\
Standard Deviation & 251.32 & 239.23 & 377.57 \\
MAD & 178.30 & 135.40 & 291.10 \\
IQR & 338.60 & 413.20 & 481.85 \\
Minimum & 15,069.40 & 15,881.20 & 15,069.40 \\
Maximum & 16,368.80 & 17,051.90 & 17,051.90 \\
Skewness & -0.26 & 0.41 & -0.04 \\
Kurtosis & 0.33 & -0.69 & -0.42 \\
\hline
\end{tabular}
\label{tab:descriptive}
\end{table}

Normality testing results indicate that pre-inauguration data generally adheres to normal distribution (4 of 5 tests indicate normality), while post-inauguration data exhibits significant non-normality (3 of 5 tests reject normality). The Shapiro-Wilk test for post-inauguration data was particularly definitive ($p = 0.0002$), confirming non-Gaussian characteristics. This finding aligns with \citet{Cont2001}, who documented that financial time series rarely follow normal distributions, especially during periods of market adjustment.

\begin{figure}[H]
\centering
\includegraphics[width=\textwidth]{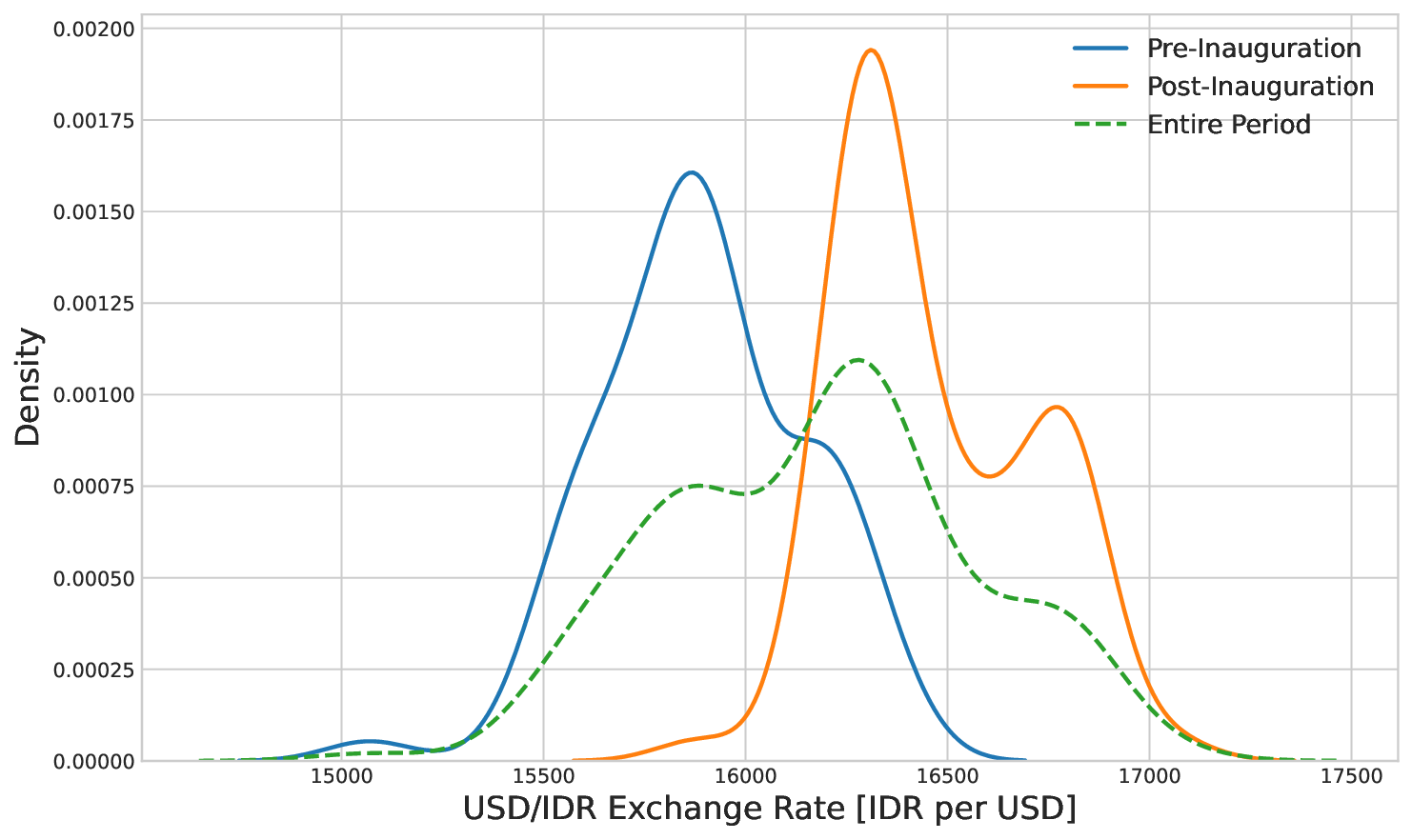}
\caption{Kernel density estimation (KDE) plots of USD/IDR exchange rates for pre-inauguration, post-inauguration, and entire period distributions. Note the clear separation and different central tendencies between pre and post-inauguration distributions.}
\label{fig:kde}
\end{figure}

The probability density distributions (Figure \ref{fig:kde}) visually confirm these distributional differences, with clear separation between pre and post-inauguration periods. The post-inauguration distribution exhibits greater positive skewness (0.41) compared to the negative skewness (-0.26) of the pre-inauguration period. This shift in distribution shape suggests greater frequency of large upward movements in the exchange rate after the political transition, consistent with the findings of \citet{Menkhoff2016}, who documented asymmetric market microstructure responses during periods of political uncertainty.

\begin{figure}[H]
\centering
\includegraphics[width=\textwidth]{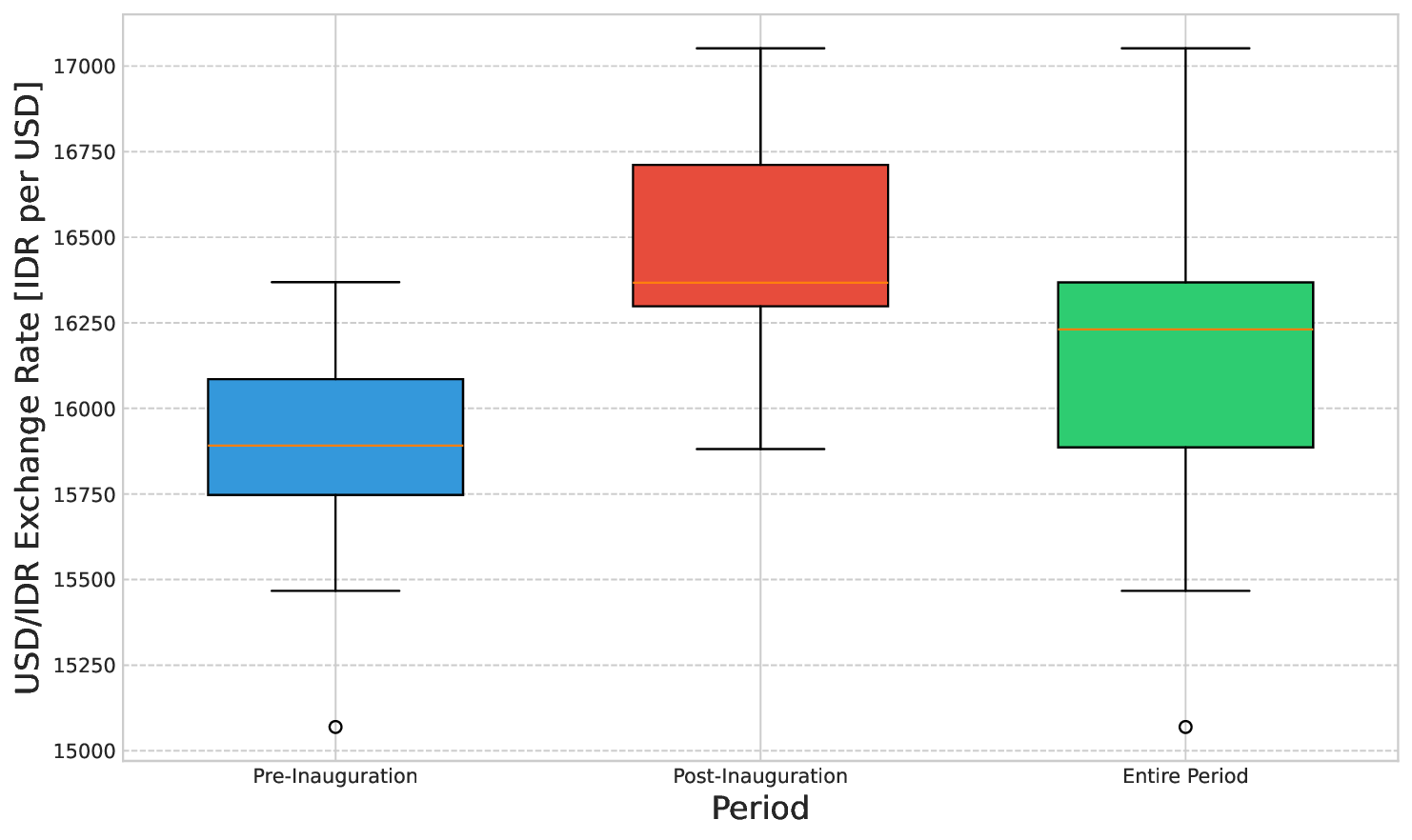}
\caption{Box plots of USD/IDR exchange rates for pre-inauguration, post-inauguration, and entire periods. The post-inauguration period shows elevated exchange rates with a similar dispersion pattern.}
\label{fig:boxplot}
\end{figure}

Box plots (Figure \ref{fig:boxplot}) further illustrate this distributional shift, demonstrating the higher median and quartile values in the post-inauguration period. The distributional changes support the policy uncertainty mechanism described by \citet{Baker2016}, reflecting the incorporation of risk premiums as market participants adjusted expectations about future US monetary, fiscal, and trade policies. Given Indonesia's significant integration with global financial markets and reliance on dollar-denominated trade, this transmission channel likely played a dominant role in the observed exchange rate dynamics.

Given the non-Gaussian characteristics of the data, we employed robust non-parametric statistical methods with bootstrap resampling ($B = 10,000$ iterations) to evaluate differences between periods. Table \ref{tab:statistics} summarizes the results of these tests, which provide strong evidence that the observed exchange rate shifts represent systematic responses to the political transition rather than random market dynamics.

\begin{table}[H]
\centering
\caption{Non-parametric Statistical Analysis of USD/IDR Exchange Rate Before and After Inauguration}
\begin{tabular}{lccccc}
\hline
\textbf{Test} & \textbf{Test Statistic} & \textbf{$p$-value} & \textbf{Bootstrap $p$} & \textbf{95\% CI} & \textbf{Effect Size} \\
\hline
Kolmogorov-Smirnov & $D = 0.841$ & $<0.0001$ & $<0.0001$ & $[0.000, 0.000]$ & $\text{RR} = 1.000^a$ \\
Mann-Whitney U & $U = 187.500$ & $<0.0001$ & $<0.0001$ & $[0.000, 0.000]$ & $\text{RR} = 1.000^a$ \\
Brown-Forsythe & $F = 0.000$ & $0.987$ & $0.504$ & $[0.025, 0.976]$ & $\text{RR} = 0.048^a$ \\
Cliff's Delta & $\delta = -0.922$ & --- & --- & $[-0.973, -0.857]$ & $\text{Large}^b$ \\
\hline
\multicolumn{6}{l}{\footnotesize{Notes: CI = Confidence Interval; RR = Rejection Ratio}} \\
\multicolumn{6}{l}{\footnotesize{$^a$ RR $>$ 0.8 indicates robust statistical significance across bootstrap samples}} \\
\multicolumn{6}{l}{\footnotesize{$^b$ Effect size interpretation based on Romano et al. (2006) thresholds: $|\delta| \geq 0.474$ indicates large effect}} \\
\end{tabular}
\label{tab:statistics}
\end{table}

The Kolmogorov-Smirnov test demonstrated highly significant differences in the cumulative distribution functions between periods ($D = 0.8414$, $p < 0.0001$), with a bootstrap rejection ratio of 1.0000. Similarly, the Mann-Whitney U test confirmed a significant difference in the stochastic ordering of the two samples ($U = 187.5000$, $p < 0.0001$, rejection ratio = 1.0000), providing further evidence that post-inauguration values are systematically higher than pre-inauguration values.

Cliff's Delta, a robust non-parametric effect size measure, revealed a large negative effect ($\delta = -0.9224$, 95\% CI: [-0.9727, -0.8571]), indicating that post-inauguration values strongly dominate pre-inauguration values. This represents a substantial effect size according to \citet{Romano2006}, who established that $|\delta| \geq 0.474$ constitutes a large effect. The robust significance across multiple non-parametric tests, combined with this large effect size, provides strong evidence that the observed shift represents a systematic response rather than random market fluctuations.

Notably, the Brown-Forsythe test for variance homogeneity showed no significant difference in volatility between periods ($F = 0.0003$, $p = 0.9873$, bootstrap $p = 0.5040$), with a variance ratio (post/pre) of 0.9061. The rejection ratio of 0.0482 provides no robust evidence of volatility differences, suggesting that while the central tendency shifted significantly, the dispersion characteristics remained relatively stable. This pattern is consistent with \citet{Gourinchas2011}, who posit that emerging markets in the "international monetary system's periphery" often react to core country political transitions through price adjustments that maintain existing volatility profiles.

The absence of significant volatility differences between periods contrasts with some historical episodes documented in previous literature. \citet{Basri2018} reported that during the 2016-2017 US presidential transition, the USD/IDR exchange rate experienced both a 7.3\% depreciation and increased volatility. The relatively stable volatility observed in our analysis may reflect the enhanced policy toolkit that \citet{Carstens2019} noted has been developed by emerging market central banks, including Bank Indonesia, to manage transition-induced market fluctuations. This suggests that while the central bank may have allowed a level adjustment in the exchange rate, it successfully contained excessive volatility through its interventions.

\begin{figure}[H]
\centering
\includegraphics[width=\textwidth]{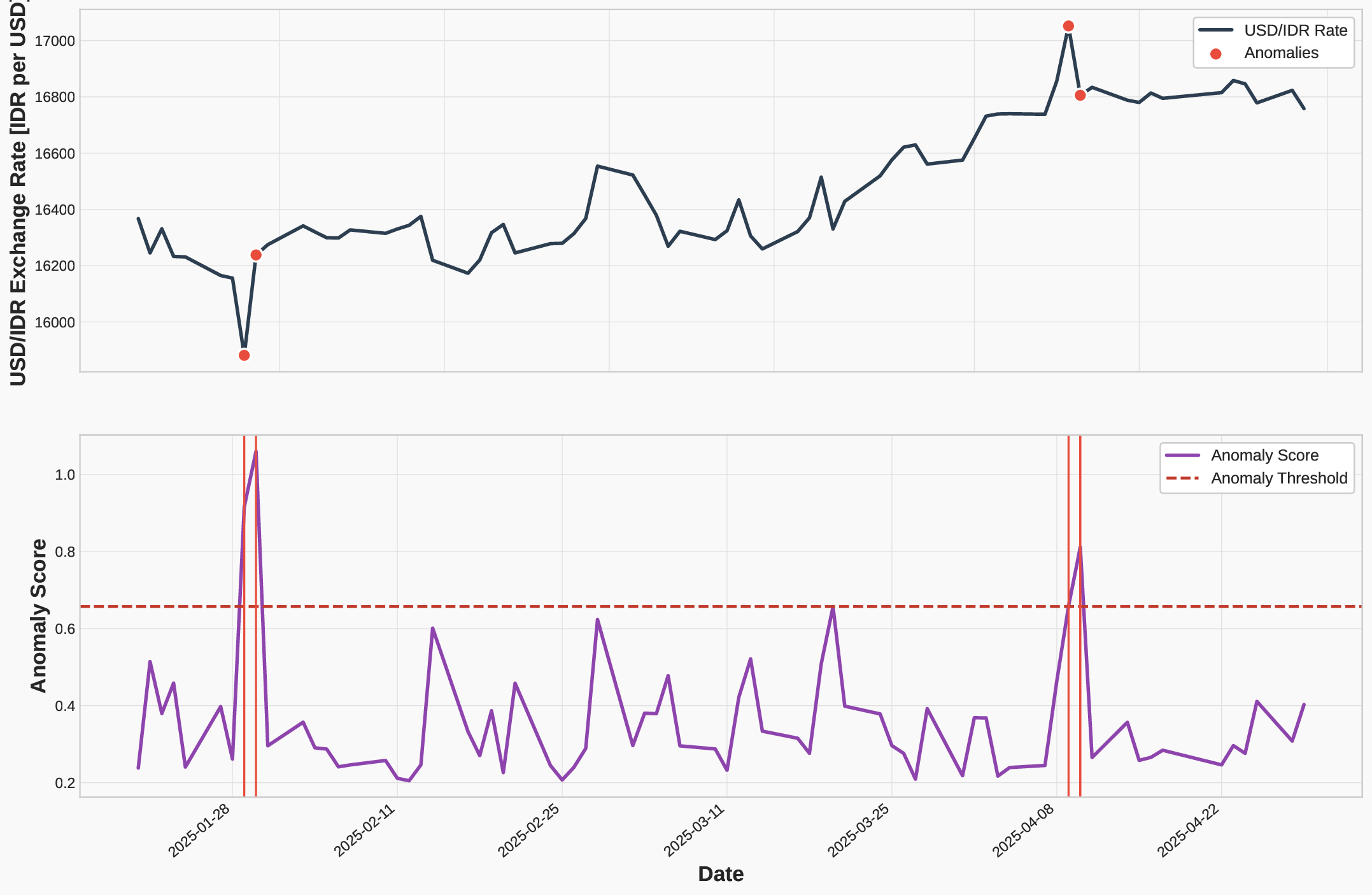}
\caption{Anomaly detection results for post-inauguration USD/IDR exchange rates. Red dots indicate detected anomalies. The bottom panel shows the anomaly score and threshold (dashed line).}
\label{fig:anomalies}
\end{figure}

Ensemble anomaly detection, combining Isolation Forest, One-Class SVM, and statistical approaches, identified four significant anomalies in the post-inauguration period, representing 5.71\% of observations (Figure \ref{fig:anomalies}). These anomalies clustered into two distinct temporal groups: late January (January 29-30, 2025) and early April (April 9-10, 2025), suggesting both immediate market reaction and delayed responses potentially triggered by specific policy announcements or implementation actions.

The most significant anomaly occurred on January 30, 2025, with an anomaly score of 1.0000 and consensus across all three detection methods. On this date, the exchange rate exhibited a sharp 2.25\% appreciation from 15,881.20 to 16,238.00, representing a significant reversal following the largest single-day depreciation in the dataset (January 29, with a -1.70\% return). The April anomalies similarly featured large movements, with April 9 showing a 1.16\% appreciation to 17,051.90 (the highest value in the dataset), followed by a sharp -1.45\% correction on April 10. The largest detected anomaly (January 30) featured a sharp correction following a significant depreciation, potentially indicating an initial overreaction followed by market reassessment. This behavior aligns with \citet{Pastor2013}, who demonstrated that policy uncertainty tends to generate initial overshooting followed by subsequent corrections as uncertainty gradually resolves.

The temporal clustering of anomalies aligns with the pattern documented by \citet{Forbes2021}, who noted that emerging market currencies often experience volatility clusters during periods of policy uncertainty, with extreme movements typically followed by sharp corrections as the market adjusts to new information.

Our findings complement the growing literature on political risk pricing in currency markets. The significant distributional shift with stable volatility aligns with the model proposed by \citet{Filippou2018}, who established that political stability measures explain time variation in currency risk premia through systematic rather than idiosyncratic channels. The large effect size quantified by Cliff's Delta indicates that this political transition constituted a major driver of exchange rate dynamics during the period under study.

The methodological approach employed in this analysis—combining multiple non-parametric tests with bootstrap resampling—provides particular confidence in the robustness of the results. As \citet{Patton2019} emphasize, this approach effectively addresses the distributional irregularities common in financial time series, especially during periods of structural change. The consistency across different test statistics, coupled with high rejection ratios, supports the conclusion that the observed exchange rate shifts represent systematic responses to the political transition rather than random market dynamics.

\section{Conclusion}

This study employed a robust non-parametric statistical framework to analyze USD/IDR exchange rate dynamics during the 100-day window surrounding the 2025 U.S. presidential inauguration. Our findings demonstrate a statistically significant and economically meaningful response in the Indonesian rupiah following the political transition. The 3.61\% depreciation observed in the post-inauguration period represents a substantial shift in currency valuation, confirmed by multiple statistical tests with consistent outcomes.

The combination of a large effect size ($\delta = -0.9224$) and robust statistical significance indicates that political transitions in major economies generate quantifiable impacts on emerging market currencies. Notably, the shift in central tendency occurred without a corresponding increase in volatility, suggesting an orderly market adjustment rather than destabilizing turbulence. This finding has important implications for monetary policy formulation, as it indicates that central bank interventions may focus more effectively on mitigating excessive volatility rather than counteracting fundamental level shifts triggered by external political events. For market participants, our results highlight the importance of incorporating political transition effects into risk management frameworks, particularly for portfolios with emerging market currency exposure. The detection of specific anomalous periods within the post-inauguration window provides valuable insights for timing-sensitive trading and hedging strategies. Additionally, the clear distributional differences observed between periods underscores the need for dynamic risk models that can adapt to regime shifts induced by major political events.

While our analysis provides robust evidence of exchange rate response patterns, several limitations warrant consideration. First, the 100-day symmetric window, though methodologically robust, may not capture longer-term adjustment processes. Second, our focus on a single currency pair limits generalizability across the broader emerging market currency landscape. Finally, isolating the specific impact of the presidential transition from concurrent economic factors presents inherent challenges. Future study could extend this statistical framework to a multi-currency panel analysis to identify systematic patterns across emerging market currencies. Additionally, investigating the relationship between specific policy announcements and detected anomalies would provide deeper insights into transmission mechanisms. Exploring interactions between exchange rate dynamics and other financial variables during political transitions represents another promising avenue for investigation.

In conclusion, our findings demonstrate that presidential transitions in the United States generate substantial, statistically significant effects on the IDR, with implications for emerging market currency dynamics more broadly. The methodological approach established in this study provides a robust framework for quantifying political risk transmission in global currency markets and highlights the importance of non-parametric techniques in financial time series analysis during periods of structural change.

\section*{Acknowledgements}
This study was supported by the UC Riverside Dean's Distinguished Fellowship in 2023. The authors would like to thank Gennady Pati from the National Development Planning Agency (Bappenas) for the valuable economic discussion. All code and data analysis used in this study are openly available on GitHub: \url{https://github.com/sandyherho/usdIDRTrump100days}.



\begin{thebibliography}{}

\bibitem[Anderson and Darling, 1952]{Anderson1952}
Anderson, T.~W. and Darling, D.~A. (1952).
\newblock {On the Distribution of the Two-Sample Cramer-von Mises Criterion}.
\newblock {\em The Annals of Mathematical Statistics}, 23(2):193--212.

\bibitem[Bailey and Chung, 1995]{Bailey1995}
Bailey, W. and Chung, Y.~P. (1995).
\newblock {Exchange Rate Fluctuations, Political Risk, and Stock Returns: Some
  Evidence from an Emerging Market}.
\newblock {\em Journal of Financial and Quantitative Analysis}, 30(4):541--561.

\bibitem[Baker et~al., 2016]{Baker2016}
Baker, S.~R., Bloom, N., and Davis, S.~J. (2016).
\newblock {Measuring Economic Policy Uncertainty}.
\newblock {\em The Quarterly Journal of Economics}, 131(4):1593--1636.

\bibitem[{Bank Indonesia}, 2023]{BankIndonesia2023}
{Bank Indonesia} (2023).
\newblock {Quarterly Report: Economic Development}.
\newblock Technical report, Bank Indonesia.

\bibitem[Basri and Hill, 2018]{Basri2018}
Basri, M.~C. and Hill, H. (2018).
\newblock {Indonesia: Twenty Years After the Asian Financial Crisis}.
\newblock {\em Bulletin of Indonesian Economic Studies}, 54(3):267--292.

\bibitem[Basri and Hill, 2020]{Basri2020}
Basri, M.~C. and Hill, H. (2020).
\newblock {The Southeast Asian Economies in the Age of Discontent}.
\newblock {\em Asian Economic Policy Review}, 15(2):185--209.

\bibitem[Bernhard and Leblang, 2002]{Bernhard2002}
Bernhard, W. and Leblang, D. (2002).
\newblock {Democratic Processes, Political Risk, and Foreign Exchange Markets}.
\newblock {\em American Journal of Political Science}, 46(2):316--333.

\bibitem[Block, 2003]{Block2003}
Block, S.~A. (2003).
\newblock {Political Conditions and Currency Crises in Emerging Markets}.
\newblock {\em Emerging Markets Review}, 4(3):287--309.

\bibitem[Brown and Forsythe, 1974]{Brown1974}
Brown, M.~B. and Forsythe, A.~B. (1974).
\newblock {Robust Tests for the Equality of Variances}.
\newblock {\em Journal of the American Statistical Association},
  69(346):364--367.

\bibitem[Carstens, 2019]{Carstens2019}
Carstens, A. (2019).
\newblock {Exchange Rates and Monetary Policy Frameworks in Emerging Market
  Economies}.
\newblock Bank for International Settlements.

\bibitem[Chohan, 2025]{Chohan2025}
Chohan, U.~W. (2025).
\newblock {Trump: A Second Administration in Seven Economic Paradoxes}.
\newblock {\em Available at SSRN 5078763}.

\bibitem[Cliff, 1993]{Cliff1993}
Cliff, N. (1993).
\newblock {Dominance Statistics: Ordinal Analyses to Answer Ordinal
  Questions}.
\newblock {\em Psychological Bulletin}, 114(3):494--509.

\bibitem[Cont, 2001]{Cont2001}
Cont, R. (2001).
\newblock {Empirical Properties of Asset Returns: Stylized Facts and
  Statistical Issues}.
\newblock {\em Quantitative Finance}, 1(2):223--236.

\bibitem[D'Agostino and Pearson, 1973]{DAgostino1973}
D'Agostino, R.~B. and Pearson, E.~S. (1973).
\newblock {Tests for Departure from Normality}.
\newblock {\em Biometrika}, 60(3):613--622.

\bibitem[Efron and Tibshirani, 1994]{Efron1994}
Efron, B. and Tibshirani, R.~J. (1994).
\newblock {\em {An Introduction to the Bootstrap}}.
\newblock Chapman and Hall/CRC.

\bibitem[Eichengreen and Gupta, 2018]{Eichengreen2018}
Eichengreen, B. and Gupta, P. (2018).
\newblock {Managing Sudden Stops}.
\newblock {\em Journal of International Economics}, 114:208--225.

\bibitem[Eshbaugh-Soha, 2005]{Eshbaugh2005}
Eshbaugh-Soha, M. (2005).
\newblock {Presidential Signaling in a Market Economy}.
\newblock {\em Presidential Studies Quarterly}, 35(4):718--735.

\bibitem[Fama, 1984]{Fama1984}
Fama, E.~F. (1984).
\newblock {Forward and Spot Exchange Rates}.
\newblock {\em Journal of Monetary Economics}, 14(3):319--338.

\bibitem[Fauvelle-Aymar and Stegmaier, 2013]{Fauvelle2013}
Fauvelle-Aymar, C. and Stegmaier, M. (2013).
\newblock {The Stock Market and U.S. Presidential Approval}.
\newblock {\em Electoral Studies}, 32(3):411--417.

\bibitem[Filippou et~al., 2018]{Filippou2018}
Filippou, I., Gozluklu, A.~E., and Taylor, M.~P. (2018).
\newblock {Global Political Risk and Currency Momentum}.
\newblock {\em Journal of Financial and Quantitative Analysis},
  53(5):2227--2259.

\bibitem[Fleming, 1962]{Fleming1962}
Fleming, J.~M. (1962).
\newblock {Domestic Financial Policies Under Fixed and Under Floating Exchange
  Rates}.
\newblock {\em IMF Economic Review}, 9:369--380.

\bibitem[Forbes and Warnock, 2021]{Forbes2021}
Forbes, K.~J. and Warnock, F.~E. (2021).
\newblock {Capital Flow Waves: Surges, Stops, Flight, and Retrenchment}.
\newblock {\em Journal of International Economics}, 88(2):235--251.

\bibitem[Gilhooly et~al., 2024]{FT2016}
Gilhooly, R., Langham, M., and Addy, T. (2024).
\newblock {What Trump 2.0 Could Mean for Emerging Markets}.
\newblock Accessed: 2025-06-02.

\bibitem[Gourinchas and Obstfeld, 2012]{Gourinchas2011}
Gourinchas, P.-O. and Obstfeld, M. (2012).
\newblock {Stories of the Twentieth Century for the Twenty-First}.
\newblock {\em American Economic Journal: Macroeconomics}, 4(1):226--265.

\bibitem[Harris et~al., 2020]{Harris2020}
Harris, C.~R., Millman, K.~J., van~der Walt, S.~J., Gommers, R., Virtanen, P.,
  Cournapeau, D., and others (2020).
\newblock {Array Programming with NumPy}.
\newblock {\em Nature}, 585:357--362.

\bibitem[Hosking, 1990]{Hosking1990}
Hosking, J. R.~M. (1990).
\newblock {L-Moments: Analysis and Estimation of Distributions Using Linear
  Combinations of Order Statistics}.
\newblock {\em Journal of the Royal Statistical Society: Series B},
  52(1):105--124.

\bibitem[{International Monetary Fund}, 2023]{IMF2023}
{International Monetary Fund} (2023).
\newblock {Indonesia: 2023 Article IV Consultation}.
\newblock Technical Report 23/145, International Monetary Fund.

\bibitem[Jarque and Bera, 1987]{Jarque1987}
Jarque, C.~M. and Bera, A.~K. (1987).
\newblock {A Test for Normality of Observations and Regression Residuals}.
\newblock {\em International Statistical Review}, 55(2):163--172.

\bibitem[Kalemli-{\"O}zcan, 2019]{Kalemli2019}
Kalemli-{\"O}zcan, S. (2019).
\newblock {U.S. Monetary Policy and International Risk Spillovers}.
\newblock {\em NBER Working Paper Series}, (26297).

\bibitem[Kolmogorov, 1933]{Kolmogorov1933}
Kolmogorov, A.~N. (1933).
\newblock {Sulla Determinazione Empirica di una Legge di Distribuzione}.
\newblock {\em Giornale dell'Istituto Italiano degli Attuari}, 4:83--91.

\bibitem[Lilliefors, 1967]{Lilliefors1967}
Lilliefors, H.~W. (1967).
\newblock {On the Kolmogorov-Smirnov Test for Normality with Mean and Variance
  Unknown}.
\newblock {\em Journal of the American Statistical Association},
  62(318):399--402.

\bibitem[Liu et~al., 2008]{Liu2008}
Liu, F.~T., Ting, K.~M., and Zhou, Z.-H. (2008).
\newblock {Isolation Forest}.
\newblock In {\em Proceedings of the 8th IEEE International Conference on Data
  Mining}, pages 413--422.

\bibitem[Mann and Whitney, 1947]{Mann1947}
Mann, H.~B. and Whitney, D.~R. (1947).
\newblock {On a Test of Whether One of Two Random Variables is Stochastically
  Larger than the Other}.
\newblock {\em The Annals of Mathematical Statistics}, 18(1):50--60.

\bibitem[McKinney, 2010]{McKinney2010}
McKinney, W. (2010).
\newblock {Data Structures for Statistical Computing in Python}.
\newblock In {\em Proceedings of the 9th Python in Science Conference}, pages
  51--56.

\bibitem[Mei and Guo, 2004]{Mei2004}
Mei, J. and Guo, L. (2004).
\newblock {Political Uncertainty, Financial Crisis and Market Volatility}.
\newblock {\em European Financial Management}, 10(4):639--657.

\bibitem[Menkhoff et~al., 2012]{Menkhoff2016}
Menkhoff, L., Sarno, L., Schmeling, M., and Schrimpf, A. (2012).
\newblock {Currency Momentum Strategies}.
\newblock {\em Journal of Financial Economics}, 106(3):660--684.

\bibitem[Mundell, 1963]{Mundell1963}
Mundell, R.~A. (1963).
\newblock {Capital Mobility and Stabilization Policy Under Fixed and Flexible
  Exchange Rates}.
\newblock {\em Canadian Journal of Economics and Political Science},
  29(4):475--485.

\bibitem[Pastor and Veronesi, 2013]{Pastor2013}
Pastor, L. and Veronesi, P. (2013).
\newblock {Political Uncertainty and Risk Premia}.
\newblock {\em Journal of Financial Economics}, 110(3):520--545.

\bibitem[Patton and Weller, 2019]{Patton2019}
Patton, A.~J. and Weller, B.~M. (2019).
\newblock {Risk Price Variation: The Missing Half of the Cross-Section of
  Expected Returns}.
\newblock {\em Review of Financial Studies}, 32(12):4687--4734.

\bibitem[Ranaraja, 2022]{Ranaraja2022}
Ranaraja, R. (2022).
\newblock {yfinance: Download Market Data from Yahoo! Finance's API}.

\bibitem[Resosudarmo and Abdurohman, 2018]{Resosudarmo2018}
Resosudarmo, B.~P. and Abdurohman (2018).
\newblock {Is Being Stuck with a Five Percent Growth Rate a New Normal for
  Indonesia?}
\newblock {\em Bulletin of Indonesian Economic Studies}, 54(2):141--164.

\bibitem[{Reuters}, 2023]{Reuters2023fall}
{Reuters} (2023).
\newblock {Indonesia C.Bank Unexpectedly Raises Rates Amid Falling Rupiah}.
\newblock Accessed: 2025-05-07.

\bibitem[{Reuters}, 2025a]{Reuters2025BI}
{Reuters} (2025a).
\newblock {Indonesia Wants Fair, Equal Relationship with US, President Says}.
\newblock Accessed: 2025-04-07.

\bibitem[{Reuters}, 2025b]{Reuters2025Bold}
{Reuters} (2025b).
\newblock {Bank Indonesia Says Will Act Boldly to Maintain Rupiah Stability}.
\newblock Accessed: 2025-05-09.

\bibitem[Rey, 2015]{Rey2015}
Rey, H. (2015).
\newblock {Dilemma not Trilemma: The Global Financial Cycle and Monetary
  Policy Independence}.
\newblock {\em NBER Working Paper Series}, (21162).

\bibitem[Romano et~al., 2006]{Romano2006}
Romano, J.~P., Shaikh, A.~M., and Wolf, M. (2006).
\newblock {A Practical Set of Threshold Values for Effect Size Measures in
  Nonparametric Inference}.
\newblock {\em Journal of Statistical Planning and Inference},
  136(12):4437--4449.

\bibitem[Rosenberger, 2024]{Rosenberger2024}
Rosenberger, L. (2024).
\newblock {Identifying Risks from US China Economic Decoupling if Trump Wins}.
\newblock In Farhadi, A., Grzegorzewski, M., and Masys, A.~J., editors, {\em
  {The Great Power Competition Volume 6: The Rise of China}}, pages 243--254.
  Springer Nature Switzerland, Cham.

\bibitem[Rousseeuw and Hubert, 2011]{Rousseeuw2011}
Rousseeuw, P.~J. and Hubert, M. (2011).
\newblock {Robust Statistics for Outlier Detection}.
\newblock {\em Wiley Interdisciplinary Reviews: Data Mining and Knowledge
  Discovery}, 1(1):73--79.

\bibitem[Sch{\"o}lkopf et~al., 2001]{Scholkopf2001}
Sch{\"o}lkopf, B., Platt, J.~C., Shawe-Taylor, J., Smola, A.~J., and
  Williamson, R.~C. (2001).
\newblock {Estimating the Support of a High-Dimensional Distribution}.
\newblock {\em Neural Computation}, 13(7):1443--1471.

\bibitem[Selmi and Bouoiyour, 2020]{Selmi2020}
Selmi, R. and Bouoiyour, J. (2020).
\newblock {The Financial Costs of Political Uncertainty: Evidence from the 2016
  US Presidential Elections}.
\newblock {\em Scottish Journal of Political Economy}, 67(2):166--185.

\bibitem[Shapiro and Wilk, 1965]{Shapiro1965}
Shapiro, S.~S. and Wilk, M.~B. (1965).
\newblock {An Analysis of Variance Test for Normality (Complete Samples)}.
\newblock {\em Biometrika}, 52(3/4):591--611.

\bibitem[Silverman, 1986]{Silverman1986}
Silverman, B.~W. (1986).
\newblock {\em {Density Estimation for Statistics and Data Analysis}}.
\newblock Chapman and Hall.

\bibitem[Smirnov, 1948]{Smirnov1948}
Smirnov, N. (1948).
\newblock {Table for Estimating the Goodness of Fit of Empirical
  Distributions}.
\newblock {\em The Annals of Mathematical Statistics}, 19(2):279--281.

\bibitem[Tabash et~al., 2024]{Tabash2024}
Tabash, M.~I., Chalissery, N., Nishad, T.~M., and Al-Absy, M. S.~M. (2024).
\newblock {Market Shocks and Stock Volatility: Evidence from Emerging and
  Developed Markets}.
\newblock {\em International Journal of Financial Studies}, 12(1):2.

\bibitem[{Trading Economics}, 2023]{TradingEconomics2023}
{Trading Economics} (2023).
\newblock {Indonesia Interest Rate}.
\newblock Trading Economics.

\bibitem[Virtanen et~al., 2020]{Virtanen2020}
Virtanen, P., Gommers, R., Oliphant, T.~E., Haberland, M., Reddy, T.,
  Cournapeau, D., and others (2020).
\newblock {SciPy 1.0: Fundamental Algorithms for Scientific Computing in
  Python}.
\newblock {\em Nature Methods}, 17:261--272.

\bibitem[Warjiyo and Juhro, 2019]{Warjiyo2019}
Warjiyo, P. and Juhro, S.~M. (2019).
\newblock {\em {Central Bank Policy: Theory and Practice}}.
\newblock Emerald Publishing, Bingley, UK.

\bibitem[Wilcox, 2012]{Wilcox2012}
Wilcox, R.~R. (2012).
\newblock {\em {Modern Statistics for the Social and Behavioral Sciences: A
  Practical Introduction}}.
\newblock CRC Press.

\bibitem[{World Bank}, 2023]{WorldBank2023}
{World Bank} (2023).
\newblock {Indonesia Economic Prospects: Navigating a Recovery}.
\newblock Technical report, World Bank.

\end{thebibliography}
\end{document}